\documentclass[final]{svjour3}
\usepackage{graphicx}
\usepackage{rotating}
\usepackage{amssymb}
\usepackage{mathptmx}
\usepackage[square,sort,comma,numbers]{natbib}
\usepackage[colorlinks=true, allcolors=blue]{hyperref}

\makeatletter
\journalname{Journal of Low Temperature Physics}

\usepackage{color}

\begin{document}

\newcommand{\redbf}[1]{\textbf{\textcolor{red}{#1}}}
\newcommand{\sups}[1]{\textsuperscript{#1}}
\newcommand{\hdblarrow}{H\makebox[0.9ex][l]{$\downdownarrows$}-}
\newcommand{\micron}{$\mu$m\xspace}

\title{Parallel Plate Capacitor Aluminum KIDs for Future Far-Infrared Space-Based Observatories}

\author{N.~F. Cothard\sups{1} \and
C.~Albert\sups{2} \and
A.~D.~Beyer\sups{3} \and
C.~M.~Bradford\sups{3,2} \and
P.~Echternach\sups{3} \and
B.~H. Eom\sups{3} \and
L.~Foote\sups{2} \and
M.~Foote\sups{3} \and
S.~Hailey-Dunsheath\sups{2} \and
R.~M.~J. Janssen\sups{3,2} \and
E.~Kane\sups{3} \and
H.~LeDuc\sups{3} \and
J.~Perido\sups{4} \and
J.~Glenn\sups{1} \and
P.~K. Day\sups{3}}

\institute{
\sups{1}NASA Goddard Space Flight Center, Greenbelt, MD 20771, USA\\
\sups{2}California Institute of Technology, Pasadena, CA 91125, USA\\
\sups{3}Jet Propulsion Laboratory, California Institute of Technology, Pasadena, CA 91109, USA\\
\sups{4}University of Colorado, Boulder, CO 80305, USA\\
\email{nicholas.f.cothard@nasa.gov}}

\maketitle

\begin{abstract}
Future space-based far-infrared astrophysical observatories will require exquis-itely sensitive detectors consistent with the low optical backgrounds. 
The PRobe far-Infrared Mission for Astrophysics (PRIMA) will deploy arrays of thousands of superconducting kinetic inductance detectors (KIDs) sensitive to radiation between 25 and 265 $\mu$m. 
Here, we present laboratory characterization of prototype, 25 -- 80 $\mu$m wavelength, low-volume, aluminum KIDs designed for the low-background environment expected with PRIMA. 
A compact parallel plate capacitor is used to minimize the detector footprint and suppress TLS noise. 
A novel resonant absorber is designed to enhance response in the band of interest. 
We present noise and optical efficiency measurements of these detectors taken with a low-background cryostat and a cryogenic blackbody. 
A microlens-hybridized KID array is found to be photon noise limited down to about 50 aW with a limiting detector NEP of about $6.5\times 10^{-19}~\textrm{W/Hz}^{1/2}$.
A fit to an NEP model shows that our optical system is well characterized and understood down to 50 aW.
We discuss future plans for low-volume aluminum KID array development as well as the testbeds used for these measurements.

\end{abstract}

\section{Introduction}
\label{sec:intro}

The PRobe far-Infrared Mission for Astrophysics (PRIMA\footnote{See \url{https://prima.ipac.caltech.edu/} for more information on PRIMA.}) is a NASA concept mission designed to address the far-infrared observatory needs laid out by the 2020 Astrophysics Decadal Survey \cite{glenn_galaxy_2021,decadal_survey_on_astronomy_and_astrophysics_2020_astro2020_pathways_2021}.
PRIMA is optimized to bridge the gap of wavelength coverage between the Atacama Large Millimeter Array and the James Webb Space Telescope and to solve a wide range of contemporary problems in astrophysics.
For example, PRIMA will enable studies of the growth of stars and black holes over cosmic time, the rise of metals and dust since cosmic noon, and the influence of cosmic magnetic fields in star and galaxy evolution.
Over its five year lifespan, PRIMA is designed to meet the objectives of the Decadal Survey and to provide the far-infrared astrophysics community with a platform to address a diverse range of scientific topics.

PRIMA will fly with broadband imaging and spectrometer instruments.
For wavelengths between 25 and 80 $\mu$m, PRIMA's imaging module (PRIMAger), will provide hyperspectral imaging with a spectral resolution of $R=\lambda/\Delta\lambda=10$.
At longer wavelengths, 80 -- 265 $\mu$m, PRIMAger will be polarization sensitive and will have four sub-bands, each with a spectral resolution of about $R=4$.
PRIMA's spectrometer module, the Far-InfraRed Enhanced Survey Spectrometer (FIRESS), will have wavelength coverage between 25 and 240 $\mu$m and will have low and high spectral resolution modes.
In low resolution mode, FIRESS will provide a spectral resolution of about $R=130$ across the entire wavelength band.
FIRESS' high resolution mode will have a  a tunable spectral resolution, with a maximum of about $R=17,000$ at $\lambda=25~\mu$m and $R=4,400$ at $\lambda=112~\mu$m.

Both PRIMAger and FIRESS will be outfitted with multiple kilo-pixel arrays of ultra-low noise superconducting kinetic inductance detectors (KIDs).
A KID is a superconducting resonator that exhibit frequency and phase shifts when incident photons break superconducting Cooper pairs and thereby change the kinetic inductance of the resonator \cite{day_broadband_2003,zmuidzinas_superconducting_2012,hailey-dunsheath_kinetic_2021}.
KIDs are typically designed to have a high quality factors and are therefore naturally multiplexed in the frequency domain, with pixel counts in the thousands per octave of readout bandwidth.
Different KID architectures will be employed for the different instrument modules and sub-bands to suit the needs of each instrument.

In this paper, we focus on the development of parallel plate capacitor (PPC) aluminum (Al) KIDs for the short wavelength band (25 -- 80 $\mu$m) of the FIRESS instrument.
Section \ref{sec:PPC-KIDs} discusses the design of the PPC Al KID and some array-level resonator characterization.
Section \ref{sec:absorber} discusses the design of the frequency selective aluminum absorber, and presents measurements of its absorption efficiency. 
Section \ref{sec:optical_measurements} presents optical noise equivalent power (NEP) measurements and the associated response and noise characterization methods.
These measurements demonstrate that our detector design can achieve photon limited performance down to an absorbed optical loading of about 50 aW, corresponding to an optical NEP of about $6.5\times 10^{-19}~\textrm{W/Hz}^{1/2}$.
In Section \ref{sec:conclusions}, we conclude and discuss the next steps for preparing these devices for PRIMA and FIRESS.
For measurements of the 
quasiparticle lifetimes of our PPC Al KIDs, we refer the reader to \cite{kane_ltd2023}.
For optical NEP measurements of the long wavelength band FIRESS detectors, we refer the reader to \cite{foote_ltd2023}.

\section{Parallel Plate Capacitor Kinetic Inductance Detectors}
\label{sec:PPC-KIDs}

\begin{figure}[ht]
\begin{center}
\includegraphics[width=0.99\linewidth, keepaspectratio]{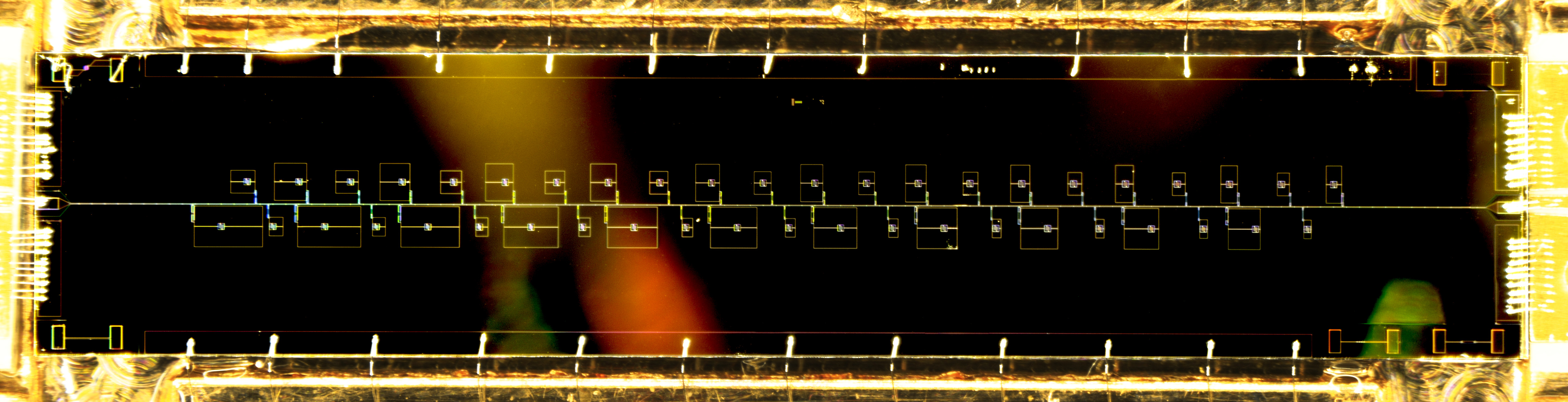}
\includegraphics[width=0.69\linewidth, keepaspectratio]{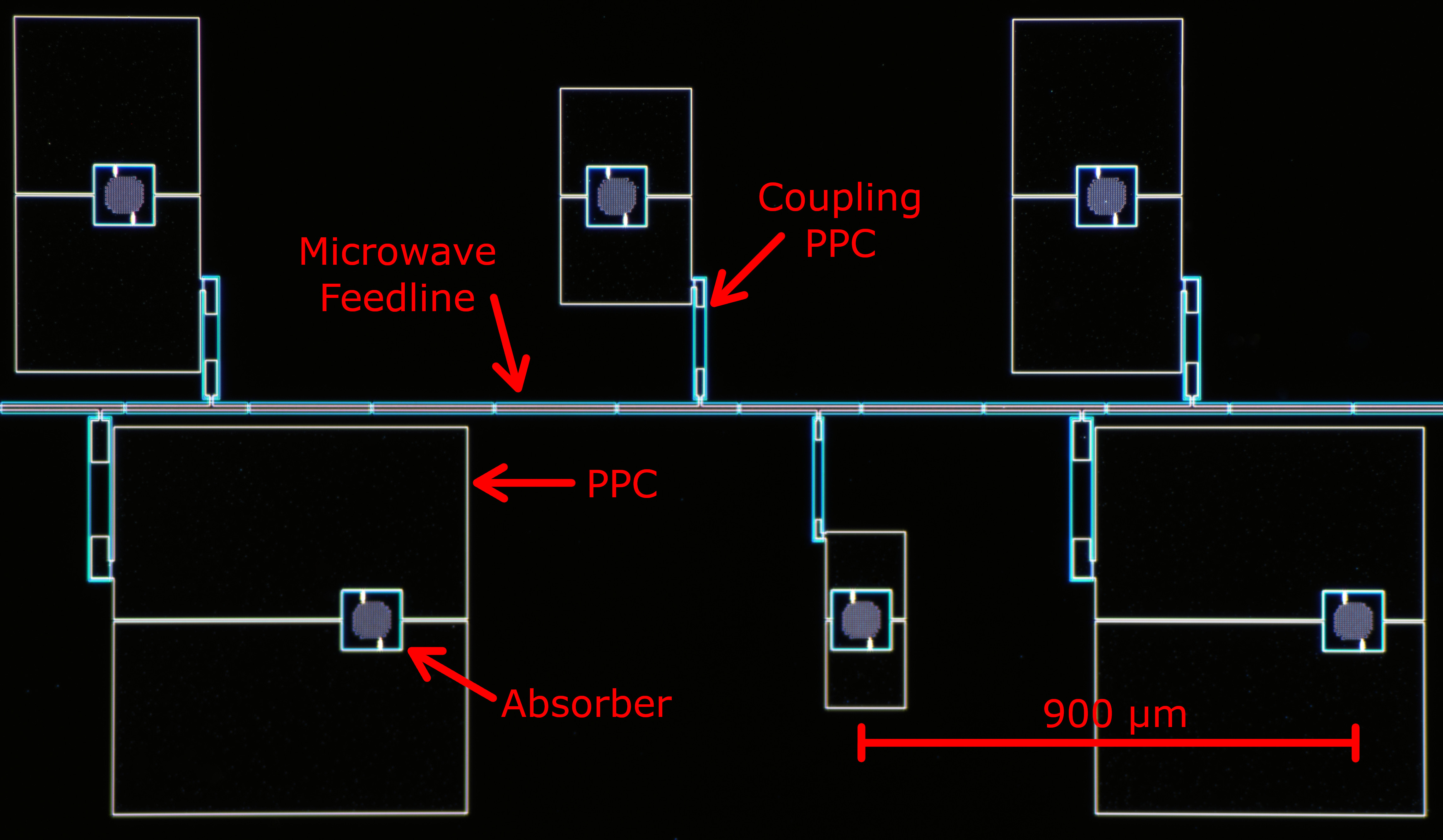}
\hfill\includegraphics[width=0.29\linewidth, keepaspectratio]{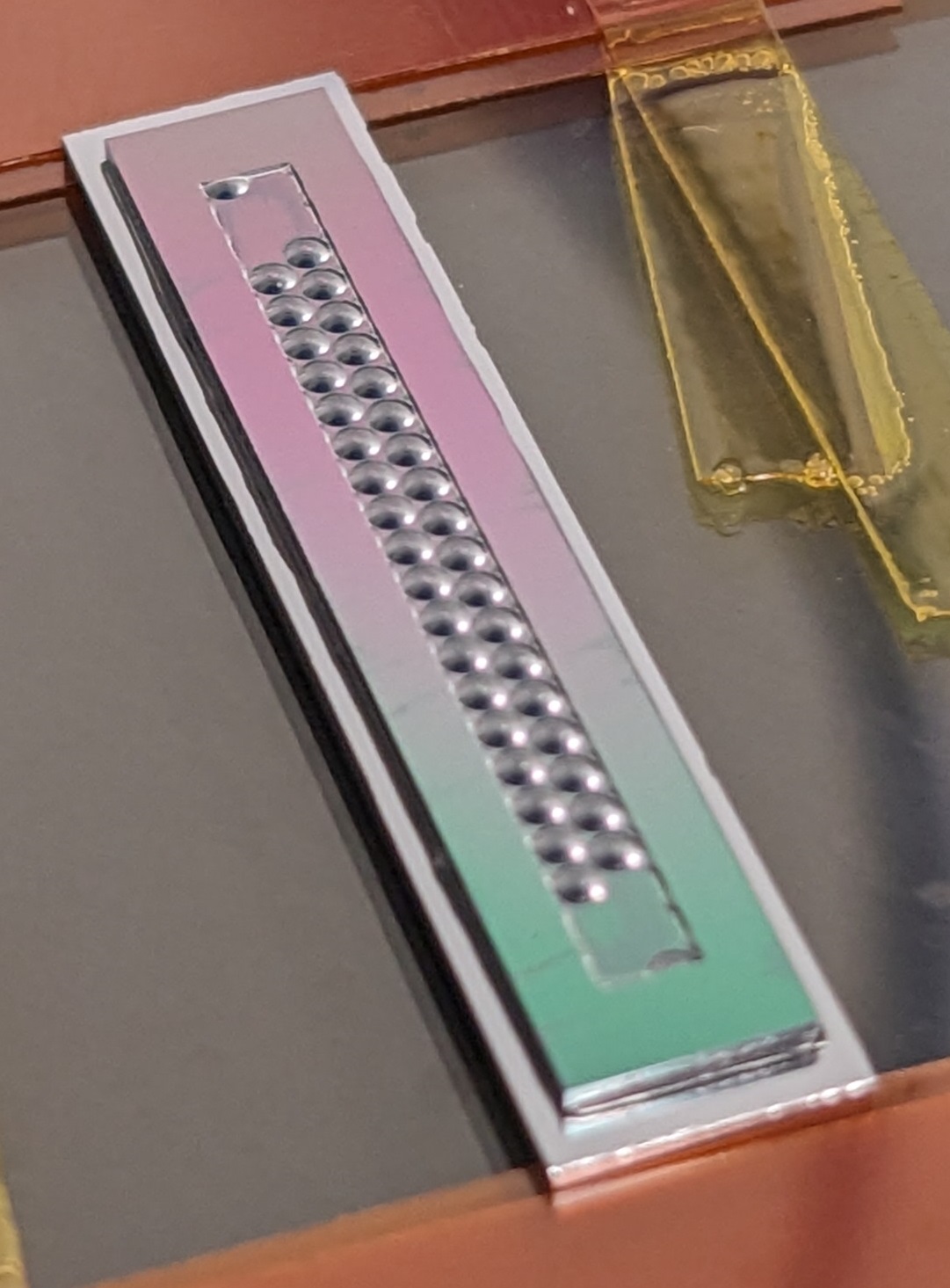}
\caption{
\textit{Top}: A photograph of a prototype 44 pixel array of PPC Al KIDs installed in a cryogenic, microwave characterization package. 
The microwave feedline runs horizontally through the center of the array, with the KIDs capacitively coupled along its length.
\textit{Bottom Left}: A microscope image of a handful of PPC Al KIDs, illustrating the pixel layout.
Aluminum absorbers are surrounded by a pair of PPCs, which are capacitively coupled to the feedline with another pair of PPCs.
The PPC's lower electrode, dielectric, and upper electrodes consist of Al, a-Si:H, and Nb, respectively.
\textit{Bottom Right}: A photograph of a prototype 44 pixel silicon microlens array that has been anti-reflection coated and hybridized to a matching Al PPC KID array to back-illuminate the detectors. 
This hybridized array was used for the optical response and NEP measurements presented in Section \ref{sec:optical_measurements}.
}
\label{fig:photos}
\end{center}
\end{figure}

The motivation for using PPCs with KIDs is to reduce the capacitor footprint, reduce the electromagnetic crosstalk between resonators from fringing fields, and to suppress TLS noise by enabling larger electric fields between electrodes.
Photographs in Figure \ref{fig:photos} illustrate the lumped-element design of our PPC Al KIDs. 
Images are shown of a prototype 44 pixel array on a 900 $\mu$m pitch that was optically tested by back illuminating the KIDs through a matching silicon microlens array, which was fabricated on a separate substrate and glued to the backside of the KID array \cite{cothard_microlenses_2023}.
For the devices presented here, the detector fabrication and optical measurements took place at NASA Jet Propulsion Laboratory (JPL).
Similarly, the microlens fabrication and the microlens-detector array hybridization took place at NASA Goddard Space Flight Center (GSFC).

Each KID consists of an absorber/inductor and a PPC, and is capacitively coupled to a single microwave feedline with another PPC.
The 70 $\mu$m diameter, 30 nm thick sputtered Al inductor/absorber consists of a meandered array of strips with resonant periodic structures, tuned to selectively absorb radiation near 25 $\mu$m wavelengths in both polarizations. 
The Al absorber design and characterization are described in detail in Section \ref{sec:absorber}.
Each end of the absorber is connected to a PPC, such that two PPCs in series form the resonator capacitor.
The lower electrode of each PPC is constructed with the same Al layer as the absorber. 
600 nm thick hydrogenated amorphous silicon (a-Si:H) is deposited via ICP-PECVD across the array, serving as the dielectric of the PPCs and as a passivation layer for the Al absorbers and the Al center feedline.
Just prior to the a-Si:H deposition (in the same vacuum), the aluminum is cleaned of oxidation using a Fluorine-based plasma etch.
The upper electrode of the PPC is a 300 nm thick layer of sputtered niobium (Nb), which also serves as the ground plane of the array. 
While the absorber of each KID is identical, the resonator capacitors are varied to set the microwave frequency of the resonator.
A second pair of smaller PPCs, constructed with the same layer stack, serve as the coupling capacitors, connecting the lumped-element KID to the feedline.
The coupling capacitor varies for each KID, tracking the resonator capacitance in order to keep a constant coupling quality factor, $Q_c \sim 40,000$ (design).
The connections between the absorber and the lower PPC electrodes are galvanically interrupted by a small strip of Nb, which traps quasiparticles inside the Al absorber due to its higher gap energy.
The Al feedline is covered with a-Si:H and has Nb ground plane cross-overs a regular intervals along its length.
Lastly, alignment marks are etched on the backside of the silicon substrate, enabling alignment with a silicon microlens die.

\begin{figure}[t]
\begin{center}
\includegraphics[width=0.99\linewidth]{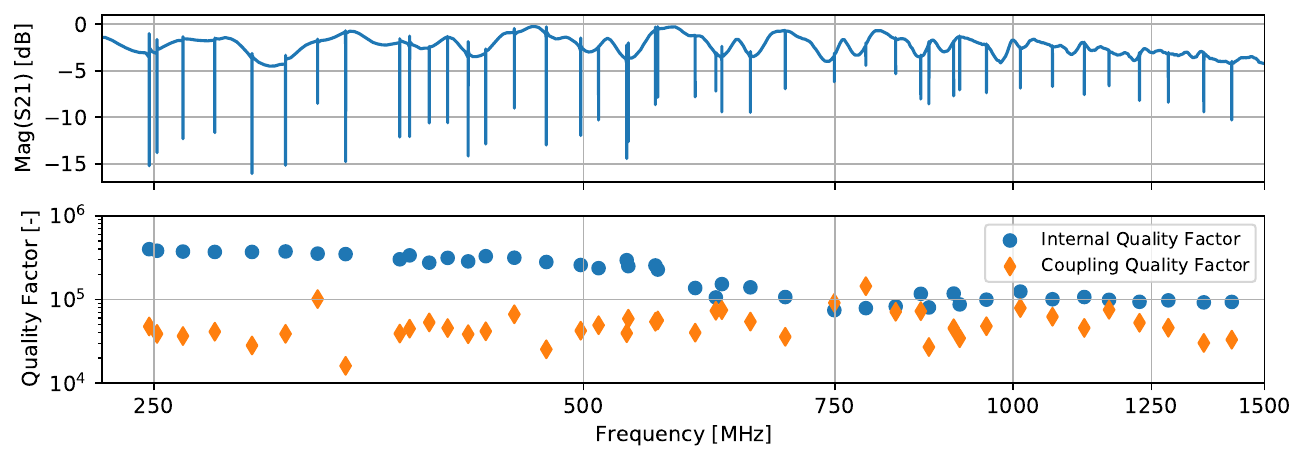}
\caption{
\textit{Top}: VNA $\textrm{S}_{21}$ measurement of prototype Al PPC KID array, showing a yield of 43 of 44 resonators.
\textit{Bottom}: Coupling and internal quality factors of the 43 yielded resonators.
The coupling quality factors, are reasonably constant with a median value fo about 45,000.
The internal quality factors are roughly $\gtrsim 100,000$.
}
\label{fig:vna-qi-qc}
\end{center}
\end{figure}

The absorber and the capacitors are designed to place the resonator frequencies in the span of 250 -- 1500 MHz with logarithmic frequency spacing.
Figure \ref{fig:vna-qi-qc} shows a measurements of the 44 pixel prototype array.
The VNA measurement indicates an detector yield of 43 resonators and that the resonant frequencies are roughly in the expected span.
A scatter plot shows the coupling and internal quality factors, $Q_i$ and $Q_c$ respectively, as a function of resonator frequency.
The median $Q_c$ is around 45,000, which is close to the expectation.
$Q_i$ decreases with frequency, indicative of TLS-limited internal quality factors \cite{zmuidzinas_superconducting_2012,pappas_two_2011}.

\section{Aluminum Far-IR Resonant Absorber}
\label{sec:absorber}

\begin{figure}[ht]
\begin{center}
\includegraphics[width=0.33\linewidth, keepaspectratio]{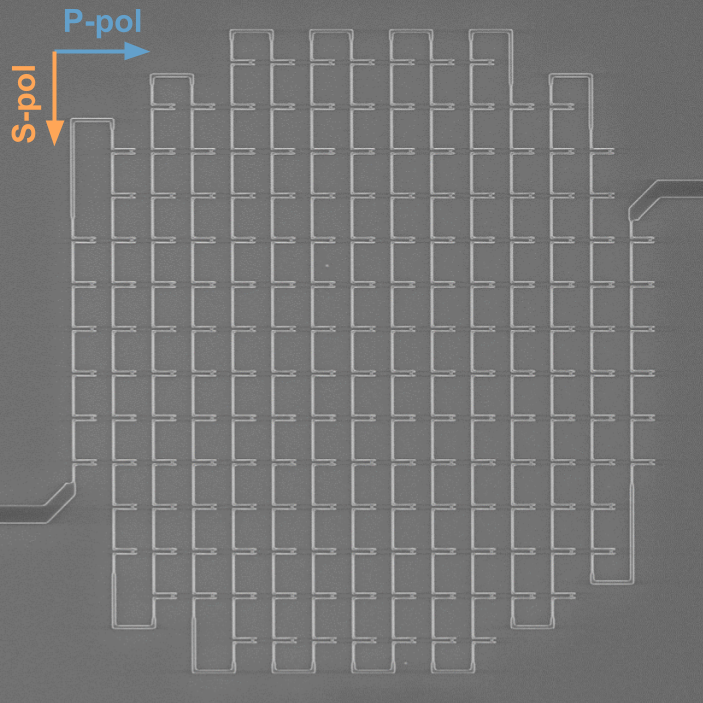}
\includegraphics[width=0.66\linewidth, keepaspectratio]{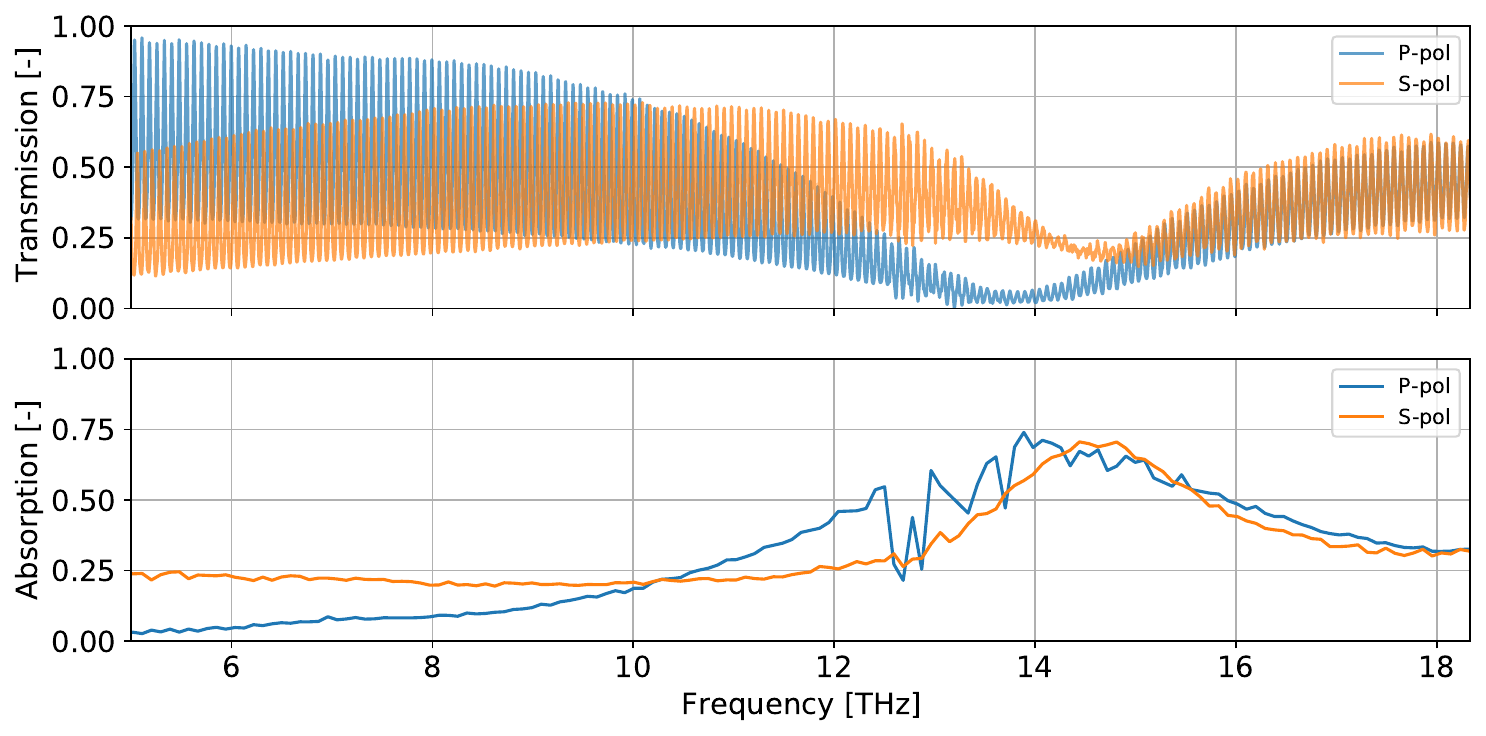}
\caption{
\textit{Left}: A scanning electron microscope image of a 70 $\mu$m diameter aluminum resonant absorber.
A vertically meandered aluminum trace is periodically interrupted by a ``hairpin'' structure, making the absorber sensitive to both polarizations.
The geometry of the hairpin structure sets the spectral width and center frequency of the absorption resonance. 
\textit{Top Right}: Polarized Fourier transform spectrometer (FTS) data of a silicon sample covered with the absorber structure, which was cooled to 5 K during measurements. 
Fabry-Perot fringing from the reflections inside the silicon wafer are visible. 
The transmission of the fringes is reduced near 12 THz (25 $\mu$m), demonstrating the frequency-selective absorption.
\textit{Bottom Right}: Absorption spectra extracted from the FTS transmission measurements.
}
\label{fig:absorber}
\end{center}
\end{figure}

For a typical mesh-grid style absorber, the low sheet resistance of aluminum films necessitates extremely narrow line widths on the order of a couple tens of nanometers. 
To work around this while maintaining high absorption efficiency of our PPC KIDs, we use a frequency selective, resonant absorber design, enabling line widths of a couple hundreds of nanometers.
As shown in Figure \ref{fig:absorber} \textit{Left}, the 70 $\mu$m diameter consists of a meandered aluminum trace. 
Periodically along the meandered path, the trace is interrupted with a notch-shaped structure that resembles a ``hairpin''. 
The geometry of the hairpin structures (including their pitch, length, and width) control the resonant frequency of the absorber and it's spectral width.
The orthogonal geometry of the hairpins with respect to the meandered path also makes the absorber sensitive in both polarizations of incident radiation.
The resonant structure is simulated and optimized prior to fabrication using Ansys HFSS. 

An initial design of the absorber was created to have a peak absorption of 25 $\mu$m. 
To test the design, a 1-inch square silicon die was patterned uniformly with the aluminum hairpin structure and meandered trace. 
The sample was cooled to 5 K and its far-infrared transmission was measured with a Fourier transform spectrometer. 
Figure \ref{fig:absorber} \textit{Top Right} shows the measured transmission in both polarizations. 
Reflections inside the silicon-substrate's cavity cause Fabry-Perot fringing, which is seen throughout the transmission spectra.
By looking at the peaks and troughs of the fringes and comparing them to those of a blank silicon sample, we estimate the absorption of the design, as shown in Figure \ref{fig:absorber} \textit{Bottom Right}.
The peak absorption has a maximum value of around 70 -- 75 \%, close to the maximum possible absorption without a backshort.
The absorption is maximized near 12 THz (25 $\mu$m), indicating that the design is working roughly as expected.

For the initial design presented in this paper, the absorption peak was found to have slightly different resonant frequencies for each polarization. 
This may be caused by differences between the patterned and designed geometry.
Further, the spectral width of the absorption resonance is slightly different in each polarization. 
The spectral width can be tuned by adjusting the Al trace width and so in a future iteration, the trace width along one direction of the pattern (corresponding to one of the polarizations) will be adjusted.
A re-optimization of the resonant absorber design is underway to match the resonant frequencies and spectral widths in both polarizations.

\section{Optical Characterization}
\label{sec:optical_measurements}

\begin{figure}[ht]
\begin{center}
\includegraphics[width=\linewidth]{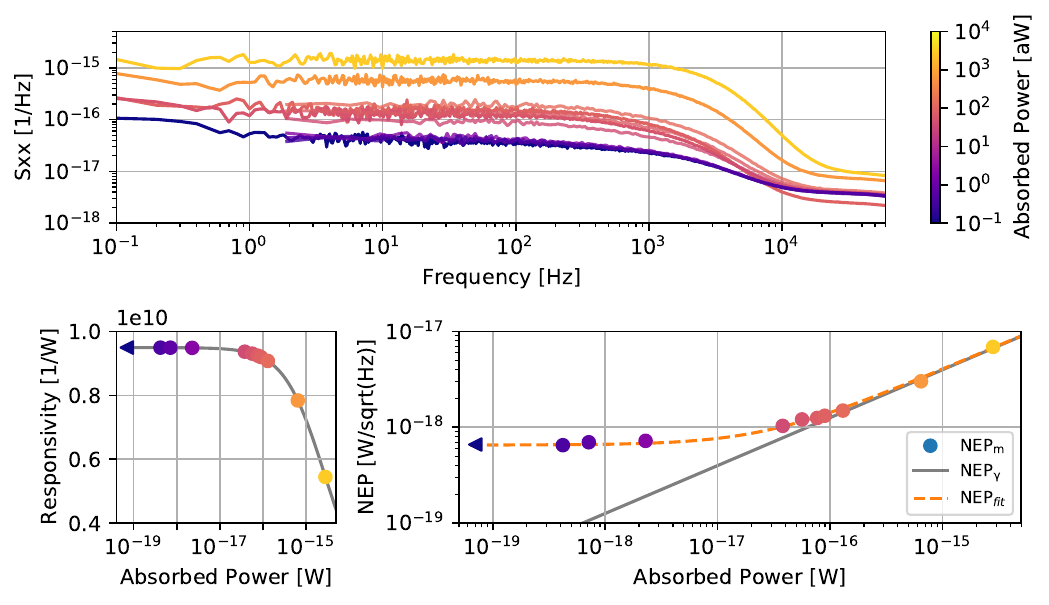}
\caption{
\textit{Top}: Fractional frequency shift noise spectra of one PPC Al KID as a function of audio frequency, which demonstrates a very white spectrum with minimal 1/f noise contributions. 
\textit{Bottom Left}: Responsivity measurements as a function of absorbed power and model fit.
\textit{Bottom Right}: Optical NEP measurement at 10 Hz as a function of absorber power compared to photon shot noise, showing photon limited performance down to about 50 aW, with a limiting NEP of about $6.5\times 10^{-19}~\textrm{W/Hz}^{1/2}$.
Data points with the $\blacktriangleleft$ symbol indicate measurements with the blackbody infrared source turned off, such that the true absorbed power of these data points is many orders of magnitude lower than plotted and are meant to illustrate the performance in the limit of zero absorbed power. 
}
\label{fig:NEP}
\end{center}
\end{figure}

A 44 pixel microlens-hybridized Al PPC KID array was optically characterized using cryogenic infrared radiation sources at JPL.
The hybridized detector array was cooled to 150 mK and housed in a light-tight box-within-a-box configuration \cite{baselmans_ultra_2012}, with a cryogenic high-permeability magnetic shield.
The detectors were illuminated through a pinhole aperture using one of two far-infrared radiation sources.
The first source was a well-characterized cryogenic blackbody that can stably achieve temperatures of up to 100 K. 
The second source was a micro-electromechanical system (MEMS) source that can be rapidly modulated from low temperature to above 200 K.
Mounted on the 3 K and 0.7 K stages of the cryostat were the pinhole aperture, a Newport ND10-10 neutral density (ND) filter and a stack of band-defining far-infrared filters from Cardiff.
The far-infrared transmission of the ND filter and Cardiff filters were cryogenically characterized with an FTS at NASA GSFC.
The bandpass of the full stack of filters is between 22 and 29 $\mu$m and the ND filter has an average transmission of 10\% within the Cardiff filter band.
The absorbed power of the detectors is calculated using the source temperature, cryogenic filter stack transmission, illumination geometry, microlens efficiency, and the absorber efficiency.

To measure the detector responsivity, VNA measurements of the resonators were taken at many far-infrared source settings. 
First, data were taken for a variety of cryogenic blackbody powers over the course of many hours.
At very low powers, a time-dependent monotonic exponentially decaying drift was found in the resonator frequencies. 
This suggested a light leak or slowly-cooling component in the cryogenic optical system.
Response measurements were then taken with the MEMS source, which was calibrated using the high power blackbody response measurements.
Figure \ref{fig:NEP} \textit{Bottom Left} shows a compilation of response measurements as a function of absorbed power. 
The measurements fit well to the model $R =  \frac{\mathrm{d} x}{\mathrm{d} P_{abs}} = A\left(1+P_{abs}/B\right)^{1/2}$ where $x=(f-f_0)/f_0$ is the fractional frequency shift of the resonator from it's zero loading resonant frequency, and $P_{abs}$ is the power absorbed in the KID. 
This model was derived from expressions for the effective quasiparticle time constant and the optical response of the resonator \cite{zmuidzinas_superconducting_2012}.

Detector noise measurements were also taken at a variety of blackbody and MEMS source powers.
For the blackbody noise measurements, the temperature of the blackbody was allowed to stabilize for a few hours before collecting data.
For the MEMS noise measurements, the MEMS source was flashed at 0.5 Hz, and a portion of the ``MEMS-on'' data was sampled, averaged, and used to compute the noise spectra.
Figure \ref{fig:NEP} \textit{Upper} shows a compilation of noise measurements in fractional frequency shift units for one resonator. 
The spectra have very low $1/f$ contributions at low frequencies.
The low frequency and TLS behavior of these and other far-infrared aluminum devices is discussed further in \cite{foote_ltd2023}.
Between 1 and 100 Hz, the spectra appear to be fairly white even at very low absorbed powers, suggesting the possibility of a light leak in the testbed.

The detector NEP is calculated using the measured noise (at 10 Hz) and responsivity, $NEP_{det}=\sqrt{S_{xx}}/R$, as a function of absorbed power.
The amplifier noise contribution, estimated at 50 kHz, is subtracted.
Figure \ref{fig:NEP} \textit{Bottom Right} shows the measured detector NEP and the theoretical photon fluctuation noise, $NEP_{\gamma}(P) = \sqrt{2Ph\nu(1/\eta_{sys}+n)}\approx\sqrt{2P_{abs}h\nu}$ where $\eta_{sys}$ is the system optical efficiency, which has been built into our calculation of $P_{abs}$, and $n$ is the photon occupation number (small for our wavelengths).
We fit to the model 
\begin{equation}
NEP_m^2 = NEP_c^2 + (NEP_\gamma(P)^2 +NEP_{GR}(P)^2)/ \eta
\end{equation}
where $NEP_c$ is a constant representing the limiting device NEP, $NEP_{GR}(P) = \sqrt{2P_{abs}\Delta/\eta_{pb}}$ is the generation-recombination NEP \cite{yates_photon_2011,hubmayr_photon-noise_2015}, and $\eta$ represents any unaccounted for optical efficiencies. 
For $NEP_{GR}$, we use a critical temperature of $T_c=1.36$ K and a quasiparticle pair breaking efficiency of $\eta_{pb}=0.13$, as measured in \cite{kane_ltd2023}.
This pair breaking efficiency was measured via near-infrared single photon events and so is a conservative lower bound.
$NEP_{GR}$ is already sub-dominant to $NEP_{shot}$ and higher values of $\eta_{pb}$ would suggest that we are closer to the photon noise limit.
The best fit model yields $NEP_c = (6.5\pm0.5)\times10^{-19}~\textrm{W/Hz}^{1/2}$ and $\eta = 1.02 \pm 0.03$.
Down to about 50 aW of absorbed power, the fit suggests that the detector is photon limited and that our system optical efficiency is understood very well.

Upcoming improvements to the detector design and the testbed design to reduce potential light leaks are expected to result in lower limiting NEPs.
Quasiparticle recombination times for this device, measured via cosmic ray pulse decay fitting, indicate time constants around 0.15 ms \cite{kane_ltd2023}.
Similar Al KIDs recently fabricated at JPL designed for 200 $\mu$m radiation time constants around 1 ms \cite{kane_ltd2023}. 
New 25 $\mu$m devices are being fabricated using methods from the 200 $\mu$m devices, which will improve the aluminum trace surface quality and increase the quasiparticle lifetime.
Increasing the lifetime from 0.15 to 1 ms could decrease the detector NEP to around $1\times 10^{-19}~\textrm{W/Hz}^{1/2}$.
Fabrication processes are being developed to reduce the trace width of the absorber, which will reduce its volume and boost its sensitivity.

\section{Conclusion and Next Steps}
\label{sec:conclusions}

We have presented the performance of prototype Al PPC KIDs with resonant absorbers. 
The successful implementation of our PPC design shows good 1/f performance and low TLS noise.
FTS transmission measurements of our initial Al resonant absorber demonstrated the effectiveness of the design and suggests a path forward for the next iteration of the design.
Optical characterization of a microlens-hybridized Al PPC KID array shows good responsivity and noise performance. 
An optical NEP of $6.5\times 10^{-19}~\textrm{W/Hz}^{1/2}$ is measured, likely limited by the short quasiparticle lifetime of this array and possible light-leaks in the optical characterization testbed.
A fit of the NEP data shows that the detector, down to at least 50 aW, is photon noise limited and that our optical system is well characterized and understood.
Fabrication of the next iteration of 25 $\mu$m detectors for PRIMA FIRESS is underway.
Improvements to the KID quasiparticle lifetime and the optical characterization testbed are expected to substantially lower the detector NEP further.

\begin{acknowledgements}
Portions of this research were carried out at the Jet Propulsion Laboratory, California Institute of Technology, under a contract with NASA (80NM0018D0004).
This work was also supported by internal research and development awards at NASA GSFC. 
NFC was supported by an NASA Postdoctoral Program Fellowship at NASA GSFC, administered by ORAU. 
JP was supported by a NASA Future Investigators in NASA Earth and Space Science and Technology Graduate Fellowship.
The authors are grateful for the support of the NASA GSFC microlens fabrication and hybridization teams.
\end{acknowledgements}

\bibliographystyle{ltd_style}
\bibliography{references2,extra}

\end{document}